\documentclass[pra,preprintnumbers,showpacs,preprint,amsmath,amssymb,floatfix]{revtex4}
\usepackage{times}
\usepackage{graphicx}
\usepackage{dcolumn}
\usepackage{bm}
\usepackage{color}
\usepackage{subfigure}
\usepackage{nicefrac}
\usepackage{mathrsfs}
\usepackage{bbm}
\usepackage{rotating}
\usepackage{textcomp}
\usepackage{fancybox}

\newcommand{\vect}[1]{\bm{#1}}
\newcommand{\imagi}{\mathbbm{i}}
\newcommand{\sech}{\text{sech}}
\newcommand{\e}{\text{e}}

\renewcommand{\tanh}{\text{tanh}}
\renewcommand{\d}{\textnormal{d}}
\renewcommand{\slash}{\not\!}
\begin{document}
\title{Nonlinear Compton scattering in ultrashort laser pulses}
\author{F. Mackenroth}
\email{felix.mackenroth@mpi-hd.mpg.de}

\author{A. Di Piazza}
\email{dipiazza@mpi-hd.mpg.de}
\affiliation{Max-Planck-Institut f\"ur Kernphysik, Saupfercheckweg 1, 69117 Heidelberg, Germany}

\date{\today}
\begin{abstract}
A detailed analysis of the photon emission spectra of an electron scattered by a laser pulse containing only very few cycles of the carrying electromagnetic field is presented. The analysis is performed in the framework of strong-field quantum electrodynamics, with the laser field taken into account exactly in the calculations. We consider different emission regimes depending on the laser intensity, placing special emphasis on the regime of one-cycle beams and of high laser intensities, where the emission spectra depend nonperturbatively on the laser intensity. In this regime, we, in particular, present an accurate stationary phase analysis of the integrals that are shown to determine the computed emission spectra. The emission spectra show significant differences with respect to those in a long pulsed or monochromatic laser field: The emission lines obtained here are much broader, and more important, no dressing of the electron mass is observed.
\end{abstract}

\maketitle

\section{Introduction}
According to classical electrodynamics an accelerated charge emits radiation. If the acceleration is due to an incident laser field this process of emission may be called inelastic scattering of the incident laser field by the charge (e.g., an electron). This process has been thoroughly investigated and is called either Thomson scattering if quantum effects are negligible \cite{Vachaspati,EberlySleeper,SarachSchapp,SalamFais,NarozhnyiFofanov2} or Compton scattering if quantum effects are important \cite{Zeldovich,NikiRitus,BaierMilstein,KibbleBrown,RitusRev,NarozhnyiFofanov1,Serbo,HarveyHeinzl}.

Throughout all of the cited works the incident laser field is taken as a plane-wave field with a peak electric field $\mathcal{E}$ and carrier angular frequency $\omega$, and its intensity is characterized by the relativistically invariant parameter
\begin{equation}
\xi = \frac{|e|\,\mathcal{E}}{\omega\,m\,c}. \label{firstxi}
\end{equation}
Here $e<0$ is the electron's charge, $m$ is its mass, and $c$ is the speed of light. A pulse in this work is indicated as \textit{intense}, if the parameter $\xi$ is much larger than unity. An electron in such an intense laser field is accelerated to relativistic velocities already in one laser period \cite{LandauII}. The electric field strengths needed to attain a parameter $\xi$ equal to unity for optical radiation and for X-rays would be $\left.\mathcal{E}\left(\hbar\omega\sim1\;\text{eV}\right)\right|_{\xi = 1} \sim 10^{10}\, \text{V}/\text{cm}$ and $\left.\mathcal{E}\left(\hbar\omega\sim1\;\text{keV}\right)\right|_{\xi = 1} \sim 10^{13}\, \text{V}/\text{cm}$, respectively. Here $\hbar$ is the usual Planck constant. These field strengths correspond to laser intensities of
\begin{eqnarray}
&\left.I\left(\hbar\omega \sim 1\; \text{eV}\right)\right|_{\xi=1} \approx 10^{18} \text{W}/\text{cm}^2\nonumber \\ &\left.I\left(\hbar\omega \sim 1 \;\text{keV}\right)\right|_{\xi=1} \approx 10^{24} \text{W}/\text{cm}^2.
\end{eqnarray}
In the optical regime ($\hbar\omega \sim 1$ eV), laser intensities of these orders have already been obtained during the last decade \cite{TajimaMourou}. Among others, these laser systems have been employed to find extensive experimental proof of nonlinear Thomson scattering \cite{EnglertRinehart,MooreKnauer,Meyerhofer,ChenUmstadter,Babzien,Kumita}. In these experiments, laser systems were employed, reaching nonlinearity parameters (this choice of language is explained below) of the order of up to $\xi \sim 1\text{-}10$. Due to lack of sufficiently intense laser systems so far, nonlinear Compton scattering in a laser field has been verified in only one experimental setup \cite{Bula}, but with the development of laser systems to ever higher peak intensities, more experimental tests for nonlinear Compton scattering seem to be in reach. The record optical intensity of $2\times 10^{22} \text{W}/\text{cm}^2$, obtained in 2008, corresponds to a parameter $\xi$ of the order of $10^2$ \cite{Yanovsky_2008}.

The parameter $\xi$ is often referred to as the nonlinearity parameter \cite{KibbleBrown,RitusRev}. In fact, for $\xi\gtrsim1$, the interaction of an electron with the laser is no longer linearly dependent on the laser intensity. This can be understood classically by observing that at electron velocities inside the laser field close to $c$, the magnetic force on the electron is of comparable strength as the electric force, and the total interaction is no longer linear in the external field because the velocity also depends on the field. Another physical interpretation for the nonlinear dependence of the scattering rates on the laser intensity can be given in the photon picture of the laser field. In fact, the intensity of a radiation field is connected to the photon number density. For a not too intense laser field, an electron will basically always scatter only one photon from it. On the other hand, if the incident radiation is very intense, that is, its photon flux is very high, the electron likely interacts with many photons from the laser field. Thus the scattering rate will no longer depend linearly on the laser intensity but will exhibit a more complex dependency. In this picture, the parameter $\xi$ gives the ratio of energy absorbed by the electron $\Delta E=|e|\mathcal{E}\lambda_{\text{C}}$ in one Compton wavelength $\lambda_{\text{C}} = \hbar/(mc)$ in units of the incident laser photon energy $E_{\omega}=\hbar\omega$. In this sense, if $\xi \gtrsim 1$, the electron on average absorbs more than one photon from the laser field during the process, which again yields nonlinear effects. So in this work, the terms \textit{multiphoton} and \textit{nonlinear} Compton scattering are used interchangeably.

The photon flux in state-of-the-art laser facilities may now become so high that the electron absorbs a large number of photons, resulting in the emission of a single high-energetic photon. If the energy of this photon is of the order of the electron's energy, then the recoil effect on the electron motion has to be taken into account. This is an intrinsically quantum effect. The strength of nonlinear quantum effects is described by the dimensionless parameter $\chi = |e|\hbar\sqrt{-\left(F_{\mu\nu}p^{\nu}\right)^2}/\left(c^3m^3\right)$, where $p^{\mu}=(\epsilon,\bm{p}c)=(m\gamma c^2,\bm{p}c)$ is the electron's initial four-momentum, where $F_{\mu\nu}=k_{\mu}a_{\nu}-k_{\nu}a_{\mu}$ is the electromagnetic field strength tensor [$k^{\mu}$ is the laser photon four-momentum and $a^{\mu}=(0,\bm{a})$ is the laser four-potential strength, with $a^2=-m^2\xi^2/e^2$] and the metric $g^{\mu\nu}=\text{diag}\left(+1,-1,-1,-1\right)$ is used in this article. It is always possible to consider the scattering of an electron by a plane wave in a reference frame in which the electron and the laser field are initially counterpropagating, and in that frame the expression of $\chi$ simplifies to
\begin{equation}
\chi = \frac{(kp)}{mc\omega}\frac{\mathcal{E}}{\mathcal{E}_{\text{cr}}}= \frac{\mathcal{E}_{\text{r.f.}}}{\mathcal{E}_{\text{cr}}}\label{chieq},
\end{equation}
where for two four-vectors $a^{\mu}$ and $b^{\mu}$ we introduced the notation $(ab)=a_{\mu}b^{\mu}$, and $\mathcal{E}_{\text{cr}} = m^2\,c^3/\hbar|e|$ is the critical field of QED. This allows the interpretation of $\chi$ as the laser's electric field amplitude evaluated in the reference frame in which the electron initially is at rest, in units of $\mathcal{E}_{\text{cr}}$. The critical field transfers an energy $\Delta E = mc^2$ to an electron over one electron Compton wavelength. Creating an electric field of that amplitude would demand a laser intensity of
\begin{equation}
I_{\text{cr}} = \frac{c}{8\pi}E_{\text{cr}}^2 = 2.3\times10^{29} \frac{\text{W}}{\text{cm}^2}. \label{critint}
\end{equation}
All laser fields available today fall short of reaching the critical field strength by at least 3 or 4 orders of magnitude. The physical relevance of this quantity is that in a constant and uniform electric field of the critical field strength, the probability of creating electron-positron pairs from a vacuum becomes non negligible \cite{HeisenbergEuler}. So if the parameter $\chi$ approaches unity, the electron will feel an electric field strength at which there are nonlinear QED effects expected to happen. It has been pointed out that for small intensity parameters $\xi \ll 1$, quantum effects scale with the laser  photon energy (in the initial rest frame of the electron), that is, with the parameter
\begin{equation}
\varrho=\frac{\hbar(kp)}{m^2c^2},
\end{equation}
while for the opposite case $\xi \gg 1$, they scale with the parameter $\chi$, that is, rather with the electric field strength \cite{RitusRev}. So in the case $\xi \gg 1$, which we put particular emphasis on, we expect for $\chi\sim1$, quantum effects to become important.

Most of the theoretical works done so far on nonlinear Compton scattering considered a monochromatic laser wave \cite{KibbleBrown,NikiRitus,BaierMilstein,RitusRev,Serbo,HarveyHeinzl}. In fact, there has been some work on electron scattering from a laser pulse of duration $\tau$ and frequency $\omega$ \cite{NarozhnyiFofanov1}, but there the authors considered a pulse fulfilling the condition
\begin{equation}
 \tau\, \omega \gg 1, \label{longpulsecond}
\end{equation}
that is, a pulse containing many cycles of the carrier field. However, in order to generate high laser peak intensities and correspondingly high $\xi$ parameters, laser pulses are compressed spatially as well as temporally. Spatial compression is usually neglected in theoretical works, assuming to deal with laser beams which are not tightly focused and which can be safely approximated by a plane wave. This is also the case of the present article. Concerning temporal compression, there have been works on nonlinear Compton scattering not relying on the condition (\ref{longpulsecond}) \cite{BocaFlor,SeiptKaempfer,HeinzlSeipt}. The resulting structure of the general scattering matrix element obtained by the authors agrees with the one found in this article. On the other hand, in \cite{BocaFlor,SeiptKaempfer,HeinzlSeipt}, the authors perform a more exploratory analysis of the scattering process in the regime $\xi\lesssim 1$, while in this article we present a detailed analytical analysis of high-intensity ($\xi \gg1$) nonlinear Compton scattering. In this work, next to the analytical part, particular emphasis is put on the effect on the spectra of varying the laser intensity and the incident electron energy. Also, in \cite{BocaFlor}, the authors consider effects of the absolute phase [the so-called carrier envelope phase (CEP)] of few-cycle pulses and its impact on the emitted photon energy spectra. Here we also consider the effect of CEP on the angular distribution of the emitted radiation (see also \cite{CEP_PRL}).

Laser pulses lasting only one or two cycles, as investigated in the present work, have become available in different frequency ranges such as in the mid-infrared \cite{Bonvalet}, in the near-infrared \cite{Krauss_2010}, in the optical \cite{Cavalieri}, and in the extreme ultraviolet regimes \cite{Sansone,Goulielmakis}. Moreover, we point out that all high-field laser facilities operating or under construction employ short pulse durations to generate high field strengths. For instance, the Petawatt field synthesizer (PFS) laser system under construction in Garching (Germany) aims at optical laser intensities of the order of $10^{22}\;\text{W/cm$^2$}$ ($\xi\approx 10^2$) by compressing an energy of $5\;\text{J}$ to only $5\;\text{fs}$, corresponding to less than two laser cycles \cite{PFS}. At the Extreme Light Infrastructure \cite{ELI} and at the High Power Laser Energy Research \cite{HiPER} facilities, which aim at unprecedented laser intensities of the order of $10^{25}\text{-}10^{26}\;\text{W/cm$^2$}$, pulse durations of about $10\;\text{fs}$ are envisaged. This shows the close linking between the generation of large values of $\xi$ and short pulse durations. Additionally, since all high-field facilities referenced here operate at optical wavelengths, we will also focus on the energy regime $\omega\sim1$ eV for the incident laser. Put in quantitative terms, short pulses containing one or only a few cycles of the electric field will be distinguished by the condition
\begin{equation}
 \tau\, \omega \sim 1, \label{shortpulsecond}
\end{equation}
in contrast to Eq.\ (\ref{longpulsecond}). For optical lasers, this corresponds to pulse durations on the order of $\tau \approx 5$ fs. We will label laser pulses fulfilling the condition (\ref{shortpulsecond}) as ultrashort. As it will turn out later, the connection between few-cycle pulses and nonlinear Compton scattering is twofold. Not only does one have to incorporate Eq.\ (\ref{shortpulsecond}) into the framework of Compton scattering, but also, nonlinear Compton scattering offers a thorough way of determining the precise temporal shape of few-cycle laser pulses \cite{CEP_PRL}.

\begin{figure}[h]
 \centering
 \includegraphics[width=\linewidth]{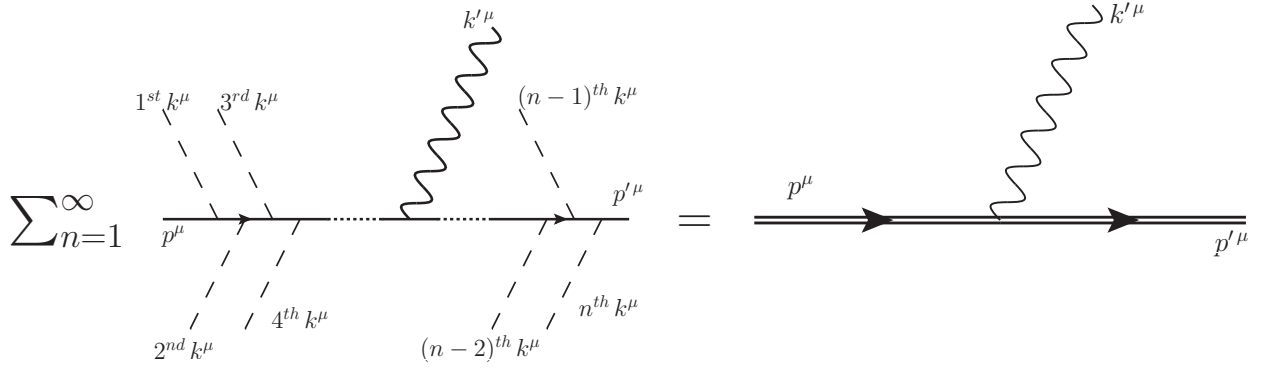}
 \caption{Feynman diagrams of multiphoton Compton scattering drawn in a conventional QED picture (left) and  in the Furry picture (right).}
 \label{feyngraph}
\end{figure}

The main purpose of this article, however, is to investigate nonlinear Compton scattering in ultrashort pulses in the framework of strong-field QED. A visualization of the scattering process we are going to consider in the language of Feynman diagrams looks like as shown on the left-hand side of Fig.\ \ref{feyngraph}.
The electron enters the laser field with an incoming four-momentum $p^{\mu}=\left(\epsilon,\bm{p}c\right)$. During the interaction with the laser pulse, it may absorb or reemit $n$ photons from or into the laser's photon field, all sharing the same wave vector $k^{\mu}$, and at some point, the electron emits a single final photon with wave vector $k^{\prime\,\mu}$. In Fig.\ \ref{feyngraph}, $n$ may be an arbitrarily large natural number (the case $n=0$ is absent because energy-momentum conservation cannot be fulfilled), what we represent by the dots inserted into the electron line. After the scattering, the electron will propagate with a changed four-momentum which, after exiting the laser field, is given by $p^{\prime\,\mu}=(\epsilon',\bm{p}'c)$. In order to take into account exactly the ultra strong laser field, which cannot be treated perturbatively, one would have to sum over all $n$ in the process depicted on the left-hand side of Fig.\ \ref{feyngraph}. Instead of doing so, we perform the calculations in the Furry picture of quantum dynamics, intrinsically taking the external field into account exactly. Here the electron states employed for computing the process amplitude are obtained by solving the Dirac equation in the plane-wave field. The resulting states are known as Volkov states, and they are indicated as a double solid line in the right-hand side of Fig.\ \ref{feyngraph}. We will restrict to values of the parameter $\chi$ smaller or of the order of 1. Therefore the main contribution to radiation comes from the diagram on the right-hand side of Fig.\ \ref{feyngraph}, and two-photon emission is roughly $\alpha_{QED}=e^2/\hbar c\approx 1/137$ times smaller than the process considered here \cite{RitusRev}.

The article is organized as follows: In section \ref{theosec}, we are going to work out the theory of the considered process. We will find that the scattering amplitude is determined by three integrals of functions of the laser pulse and of the electron parameters. These integrals are explicitly evaluated in different parameters regimes. In section \ref{resultssec}, we present emission spectra of an electron scattered from two model pulses, which we choose to model different temporal shapes of a single-cycle pulse. Finally, the summarizing discussion of section \ref{conclusionssec} concludes the article. Units with $\hbar=c=1$ are used below.
%
%
\section{Theory}\label{theosec}
The multiphoton Compton scattering process diagrammatically shown in Fig.\ \ref{feyngraph} is described by the matrix element
\begin{equation}
S_{\text{fi}} = -\imagi\,e \int \bar{\psi}_{p'\sigma'}\,\gamma_{\mu}\,\psi_{p\sigma}\sqrt{4\pi}\frac{\varepsilon^{\prime\,\mu}}{\sqrt{2\omega V}} e^{\imagi\,k'_{\mu}x^{\mu}}\d^4 x. \label{matrixelement}
\end{equation}
Here $\gamma^{\mu}$ are the usual Dirac matrices, the functions $\psi_{\bm{p}\sigma}$ and $\bar{\psi}_{\bm{p}\sigma}$ are the spinor wave functions of the electron in the background plane-wave field and its Dirac conjugate, respectively, $V$ is a normalization volume and $\mathbbm{i}$ is the imaginary unit. The electron has four-momenta $p^{\mu}$ and $p^{\prime\,\mu}$ before and after scattering, respectively. The four-vector $\varepsilon^{\prime\,\mu}$ gives the emitted photon's polarization, while $k'^{\mu} = (\omega',\bm{k}')$ is its four-momentum.

The states $\psi_{\bm{p}\sigma}$ are found as solutions of the Dirac equation
\begin{equation}\left\{\gamma_{\mu}\left[\imagi\partial^{\mu}-eA^{\mu}(\phi)\right]-m\right\} \psi_{p\sigma} = 0,\end{equation}
where the vector potential $A^{\mu}(\phi)$ describes the background plane wave and depends only on the phase $\phi=kx$, with $k^{\mu}$ being the wave vector of the laser field. The solutions of the above equation were found by Volkov already in 1935 \cite{Volkovsolution} and can be found, for example, in \cite{LandauIV}.
\begin{eqnarray}
\psi_{p\sigma} (x)&=&\frac{1}{\sqrt{2\epsilon\,V}}\left[1+\frac{e\slash k\slash \! A}{2 (kp)}\right]\,u_{p\sigma}\, \e^{\imagi\, S}, \label{volkov}
\end{eqnarray}
with the classical action
\begin{equation}
S =- (px) -\int_{-\infty}^{\phi}\d\phi' \left[ e \frac{[pA(\phi')]}{(kp)}-\frac{e^2 A^2(\phi')}{2\, (kp)}\right]
\end{equation}
and a free electron spinor $u_{p\sigma}$. The Feynman slash notation $/\! \! \!a  =\gamma_{\mu}a^{\mu}$ is used as throughout this article. For the vector potential $A^{\mu}(\phi)$ we choose the gauge in which $A^{\mu}=\psi_{\mathcal{A}}(\phi)a^{\mu}$ with $a^{\mu}=\left(0,\mathcal{A}\bm{n}\right)$, where the shape function $\psi_{\mathcal{A}}(\phi)$ gives the temporal shape of the pulse and $\bm{n}$ is a unit vector pointing along the laser's polarization axis. The electric field of the wave has an amplitude $\mathcal{E}= \mathcal{A}\omega$ and can be written as $\vect{\mathcal{E}}(\phi)=\mathcal{E}\psi_{\mathcal{E}}(\phi)\,\bm{n}$, with the electric field's shape function $\psi_{\mathcal{E}}=-\partial_{\phi}\psi_{\mathcal{A}}(\phi)$.

\begin{figure}[h]
 \centering
 \includegraphics[width=\linewidth]{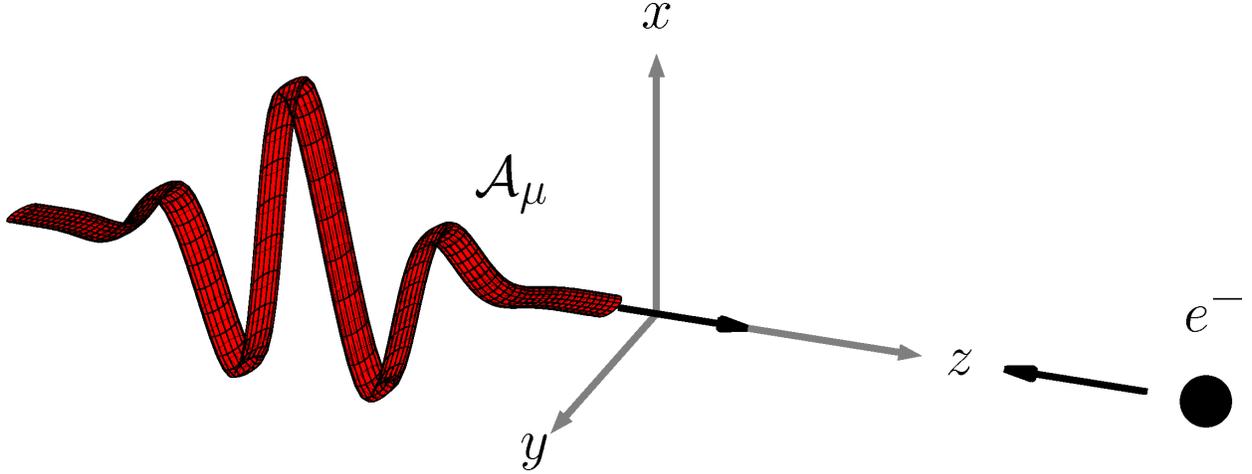}
 \caption{(Color online) General setup of the considered process.}
 \label{coordsys}
\end{figure}

We are going to investigate the scattering process in a coordinate frame where the laser pulse propagates along the positive $z$ axis, that is, is described by the wave vector $k^{\mu}=\omega(1,0,0,1)$, and where the electron before the scattering process propagates along the negative $z$ axis with the initial four-momentum $p^{\mu}=(\epsilon,0,0,-p)$. The laser is modeled to be linearly polarized along the $x$ direction. This choice of coordinates is visualized in Fig.\ \ref{coordsys}.

\begin{figure}[h]
\centering
\includegraphics[width=0.4\linewidth]{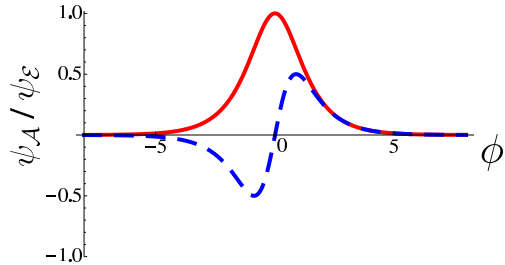}
\caption{(Color online) Shape functions according to Eq.\ (\ref{fourpotential}) $\psi_{\mathcal{A}}$ (solid red line) and $\psi_{\mathcal{E}}$ (dashed blue line).}
\label{sechfieldpic}
\end{figure}

Whenever a particular shape function is needed to obtain definite results for emission spectra, we will model the laser's four-potential by a specific choice. One possible choice used in this work is
\begin{equation}
\psi_{\mathcal{A}}=\sech \left(\phi\right), \label{fourpotential}
\end{equation}
corresponding to the electric field's shape function (see Fig.\ \ref{sechfieldpic})
\begin{equation}
\psi_{\mathcal{E}}=-\frac{1}{\omega}\frac{\partial}{\partial t} \psi_{\mathcal{A}}(\phi)= \sech(\phi)\tanh(\phi).
\label{elfield}
\end{equation}
\\The former choice models a single-cycle pulse of the electric field corresponding in the optical domain to a pulse duration of roughly $5$ fs for $\omega=1$ eV.
\begin{figure}[h]
\centering
\includegraphics[width=0.4\linewidth]{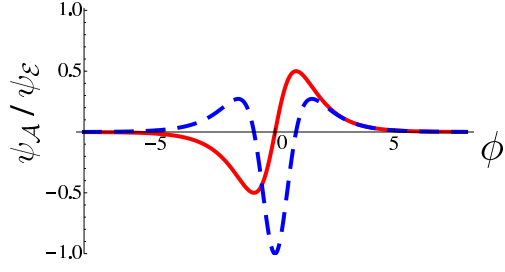}
\caption{(Color online) Shape functions according to Eq.\ (\ref{fourpotential1}) $\psi_{\mathcal{A}}$ (solid red line) and $\psi_{\mathcal{E}}$ (dashed blue line).}
\label{sechtanhfieldpic}
\end{figure}
Also, it corresponds to a sine-shaped laser pulse; that is, the maximum of the electric field's envelope coincides with a minimum of the oscillating function \cite{Krausz_2009}. To also model a cosine-shaped pulse, we consider as a second choice of four-potential
\begin{equation}
\psi_{\mathcal{A}}=\sech \left(\phi\right)\tanh(\phi), \label{fourpotential1}
\end{equation}
corresponding to the electric field's shape function
\begin{equation}
\psi_{\mathcal{E}}=\sech(\phi) - 2 \sech^3(\phi). \label{elfield1}
\end{equation}
The shape functions are show in Fig.\ \ref{sechtanhfieldpic} and the temporal pulse duration is similar to that for choice (\ref{elfield}). The above expressions of the pulse shape functions describe well a single-cycle laser pulse and have the advantage that a number of exact analytical results can be obtained.

\begin{figure}[h]
 \centering
 \includegraphics[width=0.4\linewidth]{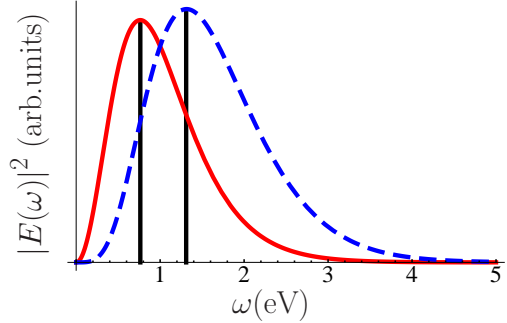}
\caption{(Color online) Electric fields in frequency space for $\psi_{\mathcal{A}}=\sech(\phi)$ (solid red line) and for $\psi_{\mathcal{A}}=\sech(\phi)\tanh(\phi)$ (dashed blue line).}
\label{fourierpic}
\end{figure}

In order to closer investigate the two choices for the shape functions (\ref{fourpotential}) and (\ref{fourpotential1}), the Fourier transforms of the two electric fields are shown in Fig.\ \ref{fourierpic}. It is clear that the two choices lead to different frequency distributions with central angular frequencies of $\omega^*_{\text{sech}}\approx0.76$ eV 
and $\omega^*_{\text{sech\,tanh}}\approx1.3$ eV, respectively.
Even though for the computation of emission spectra we need specific $\psi_{\mathcal{A}}$, many of our findings below will be valid for arbitrary shape functions.

Plugging now the solutions (\ref{volkov}) into Eq.\ (\ref{matrixelement}) we end up with the expression
\begin{equation}
S_{fi}=(2\pi)^3\delta^{(2)}\left(\bm{p}'_{\perp}+\bm{k}'_{\perp}-\bm{p}_{\perp}\right)\delta\left(\epsilon'+\omega'-\epsilon-\left(p_3'+k_3'+p\right)\right)M_{fi}, \label{finalsfi}
\end{equation}
of the matrix element $S_{fi}$, where the notation $\bm{a}_{\perp}=\left(a_x,a_y\right)$ is used and
\begin{eqnarray}
M_{fi}=-\imagi \frac{e \sqrt{2 \pi}}{\omega\sqrt{\omega'\, \epsilon'\,\epsilon\,V^3}} \bar{u}_{p'\sigma'}\,\left[\slash \varepsilon\!'^*\ f_0+e\left( \frac{\slash a\slash k\slash \varepsilon\!'^*}{2 (kp')}+\frac{\slash \varepsilon\!'^*\slash k\slash a}{2 (kp)}\right) f_1 -\frac{e^2\,a^2\ \left(k \varepsilon'^*\right) \slash k}{2(kp)(kp')}\ f_2\right]\,u_{p\sigma}, 
\end{eqnarray}
where $\varepsilon^{\prime\ast\,\mu}$ is the complex conjugate of the polarization vector of the outgoing photon and where the functions
\begin{align}
\label{f_i}
f_i=\int_{-\infty}^{\infty}\d \eta\ \psi_{\mathcal{A}}^i(\eta)
e^{\imagi\int_{-\infty}^{\eta} \d\eta'[\alpha\psi_{\mathcal{A}}(\eta')-\beta\psi_{\mathcal{A}}^2(\eta')+\zeta]},  && i\in\{0,1,2\},
\end{align}
contain all the relevant dynamical information of the process. Thus analyzing the scattering process amounts essentially to evaluating the functions (\ref{f_i}). There we have introduced the important parameters 
\begin{eqnarray}
\label{alphabetafinal}
\alpha &=& -\frac{m\xi k'_x}{(kp')}, \nonumber\\
\beta &=& -\frac{m^2\xi^2(kk')}{2(kp)(kp')},\\
\zeta &=& \omega' \frac{\epsilon + p \cos\vartheta}{(kp')}, \nonumber
\end{eqnarray}
with $\vartheta$ and $\varphi$ being the spherical angular coordinates of the emitted photon, assuming the positive $z$ axis as the polar axis.

It is convenient to introduce the notation
\begin{equation}
g(\eta)=\int_{-\infty}^{\eta} \d\eta'\ [\alpha\psi_{\mathcal{A}}(\eta')-\beta\psi_{\mathcal{A}}^2(\eta')+\zeta], \label{expfct}
\end{equation}
for the exponent in Eq.\ (\ref{f_i}). Also, we mention that the first parameter function $f_0$ is divergent because the complex exponential factor is not tending to zero for $\eta \to \pm \infty$. This problem, however, can be circumvented by expressing $f_0$ as a linear combination of the other two parameter functions, which are well defined due to the preexponential function $\psi_{\mathcal{A}}$, which goes to zero in the considered case of a pulsed laser field. In fact,
\begin{eqnarray}
\int_{-\infty}^{\infty}\d g(\eta) \e^{\imagi g(\eta)} &=& \int_{-\infty}^{0}\d g(\eta) \e^{\imagi g(\eta)} +\int_{0}^{\infty}\d g(\eta) \e^{\imagi g(\eta)}\nonumber \\
&=&\int_{-\infty}^{0}\d g(\eta) \lim_{\epsilon\to0}\e^{\imagi g(\eta)(1-\imagi\epsilon)}+\int_{0}^{\infty}\d g(\eta) \lim_{\epsilon\to0}\e^{\imagi g(\eta)(1+\imagi\epsilon)}\nonumber \\
&=&\lim_{\epsilon\to0} \left(\left.\frac{\imagi e^{\imagi g(\eta)(1-\imagi\epsilon)}}{1-\imagi \epsilon}\right|_{-\infty}^0+\left.\frac{\imagi e^{\imagi g(\eta)(1+\imagi\epsilon)}}{1+\imagi \epsilon}\right|_0^{\infty}\right)=0 \label{f0decomposition}.
\end{eqnarray}
On the other hand, it is
\begin{eqnarray}
\int_{-\infty}^{\infty}\d g(\eta) \e^{\imagi g(\eta)} &=& \int_{-\infty}^{\infty}\d \eta \frac{\d g(\eta)}{\d \eta} \e^{\imagi g(\eta)}=\alpha f_1 - \beta f_2 + \zeta f_0;
\end{eqnarray}
then
\begin{eqnarray}
f_0 &=& -\frac{\alpha f_1 - \beta f_2}{\zeta}.
\end{eqnarray}
It is interesting to note that the $\delta$ functions in Eq.\ (\ref{finalsfi}) state the energy-momentum conservation in the considered process. We have here a deep difference between the scattering of an electron off a monochromatic laser wave and off an ultrashort laser pulse. In the former case, it is possible to expand the transition matrix element (\ref{finalsfi}) into a Fourier series in which the $s^{\text{th}}$ term contains a four-dimensional energy momentum-conserving $\delta$ function of the form $\delta^{(4)}\left(sk_{\mu} + q_{\mu} - q'_{\mu}-k'_{\mu}\right)$. These $\delta$ functions can be interpreted as energy momentum conservation laws for an electron absorbing $s$ photons from the laser field and emitting a single photon of wave vector $k^{\prime\,\mu}$. The quantities $q_{\mu}$ and $q'_{\mu}$ appearing in the energy-momentum conservation laws are defined as $q_{\mu} = p_{\mu} + e^2 \mathcal{A}^2/[4(kp)]k_{\mu}$ and $q'_{\mu} = p'_{\mu} + e^2 \mathcal{A}^2/[4(kp')]k_{\mu}$ and are the so-called quasi momentum, of the electron before and after scattering (see \cite{RitusRev}). The square of this quasi momentum is equal to the square of the so-called dressed mass $m^* = m\sqrt{1+\nicefrac{\xi^2}{2}}$. From the fact that in our conservation laws, there occurs no quasi momentum, we conclude that an electron scattering off an ultrashort laser pulse inside this pulse will not behave as if it had a dressed mass, unlike in the scattering off a long pulse. This observation was just recently put into a quantitative frame considering electron-positron pair creation in ultra strong laser pulses by Heinzl et al.\ \cite{Heinzl2010}. There the authors make the interesting suggestion to recover information on the mass dressing by investigating the positions of resonance peaks in pair production spectra which are affected by an electron mass shift. From their theoretical investigations, a reduction of the effective mass inside a few-cycle pulse of an order of magnitude is obtained as compared to the monochromatic case of the same intensity. The question if an electron acquires a mass dressing could also be investigated experimentally by utilizing nonlinear Compton scattering. In fact, from the second energy momentum-conserving $\delta$ function in Eq.\ (\ref{finalsfi}), we obtain
\begin{equation}
 \omega'=\frac{\epsilon+p-(\epsilon'-p'_3)}{1-\cos(\vartheta)}. \label{ergconservation}
\end{equation}
Measuring now not only the final photon's but also the electron's momentum after scattering, this equation could be tested.

Furthermore, from Eq.\ (\ref{ergconservation}), one can easily determine the maximum energy which can be emitted at an angle $\vartheta$:
\begin{eqnarray}
\omega'_{\text{max}} = \frac{\epsilon + p}{1-\cos(\vartheta)}. \label{omgmx}
\end{eqnarray}
If, in addition to the above conservation law, we consider that $\epsilon'=\sqrt{m^2+\bm{p}'^2}$, we find an explicit formula to determine the energy of the electron after the scattering
\begin{eqnarray}
\epsilon' &=&\epsilon - \omega' + \frac{\omega' \left[\epsilon + p \cos\left(\vartheta\right)\right]}{\epsilon+p- \omega'[1-\cos\left(\vartheta\right)]}. \label{epsdelta}
\end{eqnarray}
The first energy momentum-conserving $\delta$ function in Eq.\ (\ref{finalsfi}) simply states that if the electron is initially counterpropagating with the laser field, then it holds that
\begin{equation}
 \bm{p}'_{\perp}=-\bm{k}'_{\perp}.
\end{equation}

Starting from the above $S$-matrix element $S_{fi}$, one can calculate the emitted energy spectrum $\d E/\d\omega'\d \Omega$ (average energy emitted between $\omega'$ and $\omega'+\d\omega'$, in the solid angle $\d\Omega=\sin\vartheta \d\vartheta\d\varphi$) by averaging over the initial electron spin and by summing over the final electron spin and the photon polarization. The final result reads as follows
\begin{widetext}
\begin{eqnarray}
\frac{\d E}{\d\omega'\d \Omega}&=& \frac{\omega'^2\,e^2}{\omega^2\pi^2\,\epsilon\epsilon'}\, \bigg\langle\left(\epsilon\epsilon'+p\left\{\epsilon'+\omega'\left[1-\cos(\vartheta)\right]-\epsilon- p\right\}-2m^2\right) |f_0|^2 \nonumber \\
&&-m\xi\frac{\omega^{\prime\,2}\sin(\vartheta)\left[1-\cos(\vartheta)\right]\cos\left(\varphi\right)}{\epsilon+p- \omega'\left[1-\cos(\vartheta)\right]}\text{Re}\left(f_0f_1^*\right)-m^2\,\xi^2 \text{Re}\left(f_0f_2^*\right)\nonumber \\
&&\left.+\frac{m^2\xi^2}{2}\left\{\frac{\epsilon+p- \omega'\left[1-\cos(\vartheta)\right]}{\epsilon+p}+\frac{\epsilon+p}{\epsilon+p- \omega'\left[1-\cos(\vartheta)\right]}\right\}|f_1|^2\right\rangle, \label{ergspec}
\end{eqnarray}
\end{widetext}
where the expression of $\epsilon'$ from Eq.\ (\ref{epsdelta}) has to be substituted.

The parameter functions $f_i$ given above in general are not analytically integrable. However, their properties can be investigated in limiting cases. For instance, in the low-intensity regime $\xi\ll1$, it is possible in Eq.\ (\ref{f_i}) to expand the parameter functions as perturbation series in the small parameter $\xi$, which makes them integrable. By noting that $\zeta \sim \xi^0$, $\alpha \sim \xi$, and $\beta \sim \xi^2$, and after expanding around $\xi=0$, one obtains the textbook result of single-photon Compton scattering \cite{LandauIV}.
%
%
\subsection{High intensities}
We consider here the case of a highly relativistic electron scattered by a relativistically intense laser pulse, that is, $\xi, \gamma \gg 1$. In this case, since the electron is always ultrarelativistic, one expects from classical considerations that it radiates mostly along its instantaneous velocity \cite{LandauII}, that is, close to polar angles $\pi - \vartheta \sim \xi/\gamma$ for $\gamma\gg\xi$, $\vartheta\sim \gamma/\xi$ for $\xi\gg\gamma$ or essentially within the whole interval $\vartheta\in\left[0,\pi\right]$ for $\gamma\sim\xi$. On the other hand, the expected azimuthal angle range is $\left|\varphi\right| \sim \text{min}\left\{\gamma/\xi^2,1/\gamma\right\} \ll 1$ or $\left|\pi - \varphi\right| \sim \text{min}\left\{\gamma/\xi^2,1/\gamma\right\} \ll\ 1$. As we will see, these classical considerations also hold in the quantum case. The reason is that the motion of an ultrarelativistic electron is essentially quasiclassical in the presence of under critical electromagnetic fields, and quantum effects amount to the recoil due to photon emission \cite{Baier_b_1994}. Then, at large values of $\xi$ and $\gamma$, as we are considering here, we can focus our analysis on the plane of polarization. Moreover, the above estimations on the emission region together with the fact that $\omega'\sim \text{max}\{\xi^3,\gamma^2\xi\}\omega$ \cite{LandauII} help in concluding that in the present regime, the parameters defined in Eq.\ (\ref{alphabetafinal}) scale as
\begin{eqnarray}
\alpha,\beta,\zeta \sim \xi^3. \label{abkasymp1}
\end{eqnarray}
Here it is important to note that the asymptotic relation for $\zeta$ does not hold if $1+\cos(\vartheta)\lesssim\xi^{-1}$. Thus we will have to restrict our calculations to angles $\vartheta$ not too close to $\pi$.

From these relations and from Eq.\ (\ref{f_i}), it can be seen that the exponential factor of the parameter functions is very large and the integrand is rapidly oscillating. This allows for an evaluation of the integrals by means of the stationary-phase method. The condition for finding a stationary point $\eta_0$ of the phase is given by 
\begin{eqnarray}
\alpha\psi_{\mathcal{A}}(\eta_0)-\beta\psi_{\mathcal{A}}^2(\eta_0)+\zeta \stackrel{!}{=}0, \label{statpteq}
\end{eqnarray}
which gives
\begin{eqnarray}
\psi_{\mathcal{A}}(\eta_0) &=& \frac{\alpha}{2\beta} \pm \sqrt{\left(\frac{\alpha}{2\,\beta}\right)^2 + \frac{\zeta}{2\,\beta}}.\label{statpoint}
\end{eqnarray}
From the expressions in Eq.\ (\ref{alphabetafinal}) of the parameters $\alpha$, $\beta$, and $\zeta$, it can easily be shown that the expression under the root is given by
\begin{equation}
\left(\frac{\alpha}{2\,\beta}\right)^2 + \frac{\zeta}{2\,\beta}=-\frac{1}{\left(m\,\xi\right)^2}\left[m^2+k_2^{\prime\,2}\left(\frac{\epsilon+p}{\omega'-k'_3}\right)^2\right]\equiv-\kappa^2,
\end{equation}
which is always negative and scales as $1/\xi^2\ll 1$. This fact allows in some circumstances to neglect the imaginary part of the stationary point $\eta_0$ and to treat it as a real number. Note that the above equations hold for an arbitrary shape function $\psi_{\mathcal{A}}$.

The choice of the sign in Eq.\ (\ref{statpoint}) depends on the specific shape function of the four-potential, and it has to provide that the resulting expression of the integral is not diverging. For example, we will show later that for $\psi_{\mathcal{A}}=\sech(\eta)$, we have to choose the minus sign in front of the square root. For the four-potential in Eq.\ (\ref{fourpotential1}), on the other hand, the choice of the sign depends on the specific stationary point we are considering.

Since, in the limit $\xi \to \infty$, it holds that $\alpha$ and $\beta$ are of the same order, then 
\begin{equation}
\left.\psi_{\mathcal{A}}(\eta_0)\right|_{\xi \to \infty}=\frac{\alpha}{2\, \beta} = \cot\left(\frac{\vartheta}{2}\right) \cos\left(\varphi\right)\frac{\gamma+\sqrt{\gamma^2-1}}{\xi}, \label{statpointeq}
\end{equation}
and $\eta_0$ is real only for 
\begin{equation}
\psi_{\mathcal{A},\text{min}}\leq\frac{\alpha}{2\,\beta}\leq\psi_{\mathcal{A},\text{max}}, \label{statptcond}
\end{equation} 
where $\psi_{\mathcal{A},\text{min/max}}$ are the minimum and maximum value which the function $\psi_{\mathcal{A}}(\eta)$ takes for real $\eta$, respectively. The condition (\ref{statptcond}) is important because if the stationary point has an imaginary part, the corresponding integral in Eq.\ (\ref{f_i}) contains an exponentially damping factor, which in turn implies that the emission is suppressed (note that the possibility that the integral shows an exponentially amplifying term is excluded from physical considerations). Consequently, we consider only such situations in which the condition (\ref{statptcond}) is fulfilled. Once the process parameters such as the laser intensity and the incoming electron energy are fixed according to a specific experimental setup, the condition (\ref{statptcond}) turns into a boundary condition for the observation angles $\vartheta$ and $\varphi$. In the case under consideration here, the emission will be detectable only in a narrow cone around the azimuthal angles $\varphi=0$ and $\varphi=\pi$. Thus, fixing $\varphi$ to one of these values, the condition in Eq.\ (\ref{statptcond}) provides two values $\vartheta_{\text{min}}$ and $\vartheta_{\text{max}}$, confining the polar angle range within which significant radiation is expected. Considering that for the choice (\ref{fourpotential}), we have $\psi_{\mathcal{A},\text{min}}=0$ and $\psi_{\mathcal{A},\text{max}}=1$, one, for instance, finds from Eq.\ (\ref{statptcond}) that at $\varphi=\pi$, there is no emission expected. Furthermore, it is found that to observe emission at $\vartheta_{\text{min}}\leq90^{\circ}$, it must be $\xi \geq 2\gamma$. For the choice (\ref{fourpotential1}), on the other hand, we have $\psi_{\mathcal{A},\text{min}}=-0.5$ and $\psi_{\mathcal{A},\text{max}}=0.5$, and consequently, emission into the half space $\vartheta_{\text{min}}\leq90^{\circ}$ is observed only for $\xi \geq 4\gamma$. On this respect, we observe that, based on the relation (\ref{statptcond}), we have proposed in \cite{CEP_PRL} a method for determining the CEP of laser pulses with $\xi \gg 1$.

Now, following the method of stationary phase, we expand the exponential function as well as the preexponentials in the integrals (\ref{f_i}) in a perturbation series in $\left(\eta-\eta_0\right)$ since for values of $\eta$ far away from the stationary point, the rapid oscillations of the integrand will suppress any contribution to the integral's value:
\begin{equation}
f_i \dot{=} \int_{-\infty}^{\infty}\d \eta \sum_{n=0}^{N}
\left.\frac{(\eta-\eta_0)^n}{n!}\left(\frac{\partial}{\partial \eta}\right)^n\left[\psi_{\mathcal{A}}^i(\eta)\right]\right|_{\eta=\eta_0} \exp\left[\imagi\sum_{m=0}^{M}\left.\frac{(\eta-\eta_0)^m}{m!}\left(\frac{\partial}{\partial \eta}\right)^m g(\eta)\right|_{\eta=\eta_0}\right], \label{expandedfj}
\end{equation}
with $g(\eta)$ being given in Eq.\ (\ref{expfct}) and $N$ and $M$ being the orders up to which the preexponential and the exponential functions are expanded, respectively. Due to cancellations in the squared matrix element, the preexponential perturbation series needs to be taken into account up to second order. The exponential series, on the other hand, needs to be considered up to third order. In fact, from the above expression of the function $g(\eta)$, we have
\begin{eqnarray}
g''(\eta) &=& \alpha \psi_{\mathcal{A}}' - 2 \beta\,\psi_{\mathcal{A}}'\psi_{\mathcal{A}},  \nonumber\\
g'''(\eta) &=& \alpha \psi_{\mathcal{A}}'' - 2\beta\left(\psi_{\mathcal{A}}''\psi_{\mathcal{A}}+\psi_{\mathcal{A}}^{\prime\,2}\right). \label{derivatives}
\end{eqnarray}
From Eq.\ (\ref{statpoint}) we obtain $g''(\eta_0) = \mp \imagi\,2\beta \kappa \psi_{\mathcal{A}}'\sim \xi^2$ and $g'''(\eta_0)= 2\beta\psi_{\mathcal{A}}^{\prime\,2} \sim \xi^3$. Therefore the values of $\eta$ around $\eta_0$ contributing to the integrals are such that $(\eta-\eta_0)\sim 1/\xi$ and the second- and the third-order terms in the exponential give contributions of the same order of magnitude. 

From the above considerations, Eq.\ (\ref{expandedfj}) takes the form 
\begin{eqnarray}
f_i &\sim& \sum_lG^{(0)}_{i,l} \mathcal{I}_{0,l} + G^{(1)}_{i,l} \mathcal{I}_{1,l} + \frac{1}{2} G^{(2)}_{i,l}\mathcal{I}_{2,l}, \label{gnrlasympexp}
\end{eqnarray}
where $l$ is an index running over all stationary points $\eta_{0,l}$ found from solving Eq.\ (\ref{statpointeq}). For reasons of convenience, we defined $G_{i,l}^{(0)} := G_i(\eta_{0,l})$, $G^{(1)}_{i,l} := G'_i(\eta_{0,l})$ and $G^{(2)}_{i,l} := G''_i(\eta_{0,l})$ and the integrals
\begin{align}
\mathcal{I}_{i,l} &=\int_{-\infty}^{\infty} \d \eta \ (\eta-\eta_{0,l})^i\exp\left[\imagi\left(g_{0,l}+g^{(2)}_{0,l}\frac{(\eta-\eta_{0,l})^2}{2}+g^{(3)}_{0,l}\frac{(\eta-\eta_{0,l})^3}{6}\right)\right]= \nonumber \\
&= \exp\left[\imagi\left(g_{0,l} + \frac 13 \frac{g^{(2)^3}_{0,l}}{g^{(3)^2}_{0,l}}\right)\right]\int_{-\infty}^{\infty} \d y_l \ y_l^i\exp\left[\imagi\left(\frac{g^{(3)}_{0,l}}{6}y_l^3 - \frac 12 \frac{g^{(2)^2}_{0,l}}{g^{(3)}_{0,l}}y_l\right)\right], \label{expandedfi}
\end{align}
where we have introduced the quantities $g_{0,l} := g(\eta_{0,l})$, $g^{(2)}_{0,l} := g''(\eta_{0,l})$ and $g^{(3)}_{0,l} := g'''(\eta_{0,l})$ and where $y_l = \eta - \eta_{0,l} - b_l$ with $b_l = -g^{(2)}_{0,l}/g^{(3)}_{0,l}\ll 1$. In order to work out the exponential factor outside of the integral in Eq.\ (\ref{expandedfi}), we note that
\begin{equation}
\frac{g^{(2)^3}_{0,l}}{g^{(3)^2}_{0,l}}=-\imagi \frac{2\left(\pm\kappa^3\right)\beta}{\psi_{\mathcal{A}}'(\eta_{0,l})}, \label{derivsratio}
\end{equation}
where the freedom to choose the sign of $\kappa$ is written explicitly. Thus this term gives a purely real contribution to the exponential in Eq.\ (\ref{expandedfi}). If we next expand $g_{0,l}$ as a function of $\kappa\ll 1$, around zero up to third order, we can easily show that the imaginary part of $g_{0,l}$ exactly cancels the quantity $g^{(2)^3}_{0,l}/g^{(3)^2}_{0,l}$ in such a way that
\begin{equation}
g_{0,l} + \frac 13 \frac{g^{(2)^3}_{0,l}}{g^{(3)^2}_{0,l}}\approx \text{Re}(g_{0,l}).
\end{equation}
Finally, if $i=0$ in Eq.\ (\ref{gnrlasympexp}), then the integration in $y_l$ in Eq.\ (\ref{expandedfi}) can be performed analytically, and the result is
\begin{eqnarray}
\mathcal{I}_{0,l} \sim 2 \,\e^{\imagi\text{Re}(g_{0,l})}\left(-\frac{2\,\pi^3}{g^{(3)}_{0,l}}\right)^{\frac13} \text{Ai}\left(\lambda_l\right), \label{J0final}
\end{eqnarray}
where Ai$(x)$ is the Airy function of first kind \cite{AbramSteg} and its argument $\lambda_l$ is defined as
\begin{equation}
\lambda_l = \left(\frac{2\kappa^3\beta}{\psi_{\mathcal{A}}'(\eta_{0,l})}\right)^{\frac23}. \label{airyarg}
\end{equation}
Now, from the first line in Eq.\ (\ref{expandedfi}) we obtain that 
\begin{eqnarray}
\mathcal{I}_{2,l} &=& 2\,\frac{\partial}{\partial\,g^{(2)}_{0,l}} \mathcal{I}_{0,l}\sim  4  \e^{\imagi\text{Re}(g_{0,l})}\left(\frac{2\imagi\pi^3}{g^{(3)}_{0,l} g^{(2)^3}_{0,l}}\right)^{\frac13}\!\!\!\left(\lambda_l^{\frac32}\, \text{Ai}\left(\lambda_l\right) +  \lambda_l\, \text{Ai}'\left(\lambda_l\right) \right)\!. \label{J2final}
\end{eqnarray}
Finally, in order to compute the last integral $\mathcal{I}_{1,l}$, we employ the same technique as in Eq.\ (\ref{f0decomposition}) and obtain
\begin{equation}
 \mathcal{I}_1 = -\frac{g^{(3)}_{0,l}}{2\,g^{(2)}_{0,l}} \mathcal{I}_2. \label{J1final}
\end{equation}
In conclusion, Eqs. (\ref{gnrlasympexp}), (\ref{J0final}), (\ref{J2final}) and (\ref{J1final}) provide us with the asymptotic expansion of the $f_i$ in the ultrarelativistic regime.

To illustrate the above techniques, we apply them to the shape functions (\ref{fourpotential}) and (\ref{fourpotential1}). In the former case we note that by virtue of symmetry considerations, the parameter functions $f_i$ need to be evaluated only in the interval $\eta\in\left[0,\infty\right]$. Thus Eq.\ (\ref{statpointeq}) provides as a unique stationary point the positive solution of the equation
\begin{equation}
\eta_0=\text{arcsech} \left(\frac{\alpha}{2\beta}\right).
\end{equation}
The exponential function in this case is given by
\begin{equation}
g(\eta)=2\, \alpha\, \text{arctan}\! \left[ \tanh\left(\frac{\eta}{2}\right)\right] - \beta\, \tanh(\eta)+ \gamma\,\eta, \label{expfct1}
\end{equation}
where an unimportant constant phase was dropped. According to Eqs.\ (\ref{derivatives}) and the discussion following it, the second and third derivatives of (\ref{expfct1}) at the stationary point are
\begin{eqnarray}
g^{(2)}_0 &=& \mp \imagi\,\kappa\, \frac{\alpha^2}{2\beta}\sqrt{1-\left(\frac{\alpha}{2\beta}\right)^2} \nonumber \\
g^{(3)}_0 &=& 2\beta\, \left(\frac{\alpha}{2\beta}\right)^2\left(1-\left(\frac{\alpha}{2\beta}\right)^2\right). \label{expderivatives} 
\end{eqnarray}
Since it must be $\imagi g^{(2)}_0 <0$ to obtain convergent integrals $\mathcal{I}_{i,l}$, and noting that $\beta<0$, we find that for the choice (\ref{fourpotential}) in Eq.\ (\ref{statpoint}), one has to choose the negative sign of $\kappa$. Finally, by plugging now the solutions (\ref{expderivatives}) into the above equations, we immediately find the asymptotic transition amplitude.

Next, for the cosine-shaped pulse arising from the choice (\ref{fourpotential1}), we meet a different situation. Since in this case the integrals $f_i$ do not feature symmetry properties, we have to integrate the whole range $\eta \in \left[-\infty,\infty\right]$. Then, however, Eq.\ (\ref{statpointeq}) gives two real solutions $\eta_{0,i}$
\begin{eqnarray}
\eta_{0,1} &=& \text{sgn}\left(\frac{\alpha}{2\beta}\right)\ \text{arcsech} \left[\sqrt{\frac{1+\sqrt{1-\left(\nicefrac{\alpha}{\beta}\right)^2}}{2}}\right]\nonumber \\
\eta_{0,2} &=& \text{sgn}\left(\frac{\alpha}{2\beta}\right)\ \text{arcsech} \left[\sqrt{\frac{2\left(\nicefrac{\alpha}{2\beta}\right)^2}{1+\sqrt{1-(\nicefrac{\alpha}{\beta})^2}}}\right].
\end{eqnarray}
It is easy to obtain $g^{(2)}_{0,l}$ and $g^{(3)}_{0,l}$ by plugging these latter expressions into Eqs. (\ref{derivatives}), but the resulting expressions are cumbersome and thus not reported. The asymptotic expansion of the $f_i$ is then simply obtained by summing up the contributions from the two stationary points.
%
%
\subsection{Classical limit}
In electron-laser scattering, there are essentially two types of quantum effects. First the fact that the motion of the electron is quantum. However, due to the fact that at $\xi,\gamma\gg1$, the electron is highly relativistic and its typical De Broglie wavelength is very small, this effect may be neglected. On the other hand, if an electron emits a photon whose energy is comparable to the electron's energy, it will feel a recoil. Since, for the incident laser pulse, we are considering optical frequencies, the only energy which can possibly be comparable to the electron's initial energy is the outgoing photon's energy. The classical limit is therefore defined by the condition \cite{RitusRev}
\begin{equation}\frac{(kk')}{(kp)} \ll 1.\label{classcond}\end{equation}
This allows for some major simplifications in the emission probability [Eq.\ \ref{ergspec}]:
\begin{eqnarray}
\frac{\d E}{\d\omega'\d \Omega}&\stackrel{(kk')\ll (kp)}{\approx}& \frac{\omega'^2\,e^2}{\omega^2\pi^2\,\,\gamma^2}\,
\left\{\xi^2\left[|f_1|^2-\text{Re}\left(f_0f_2^*\right)\right]-|f_0|^2\right\}, \label{classlim}
\end{eqnarray}
with the functions $f_i$ still given by Eq.\ (\ref{f_i}) but with the parameters $\alpha$, $\beta$, and $\zeta$ given by Eq.\ (\ref{alphabetafinal}) with the substitution $(kp')\to(kp)$. In this way, one exactly recovers the classical expression of the energy spectrum calculated from the trajectory of the electron in a plane wave and from the Li\'{e}nard-Wiechert potentials \cite{LandauII}.
%
%
\section{Numerical spectra}\label{resultssec}
The analytical results obtained in the previous section give the differential photon energy spectrum of nonlinear Compton scattering in ultrashort laser pulses. We are going to show low-intensity emission spectra for the choice (\ref{fourpotential}) only. For high intensities $\xi \gg1$, however, we aim to show the applicability of the stationary phase analysis to arbitrary shape functions and show emission spectra for the choice (\ref{fourpotential1}) as well. All spectra are plotted at the observation angle $\varphi = 0$, unless otherwise mentioned. This choice is justified by the fact that an ultrarelativistic electron mainly emits in the plane determined by the laser polarization and propagation direction.
%
\subsection{Low intensity regime}\label{classresultssec}
\begin{figure}[h]
\centering
\includegraphics[width=\linewidth]{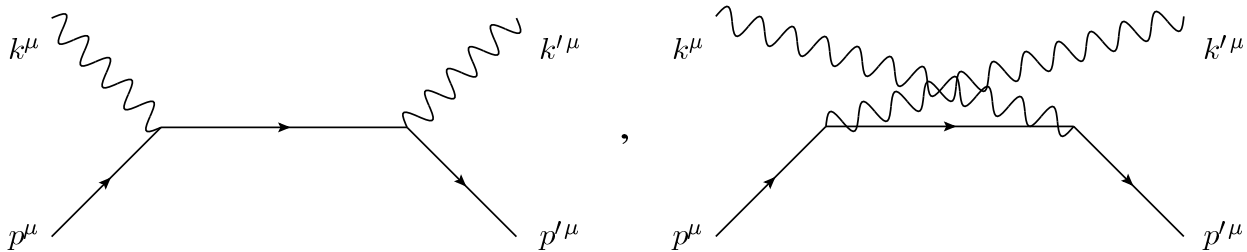}
\caption{Diagrams contributing to single-photon Compton scattering. Note that, as it is known, two diagrams contribute to the lowest order Compton scattering. By working in the Furry picture, both diagrams are automatically taken into account.}
 \label{singlecomptonpic}
\end{figure}

We characterize the low-intensity regime by either $\xi \ll 1$ or $\xi\sim1$. According to the arguments given in the introduction, the regime in which $\xi\ll 1$ corresponds to scattering dominated by the single-photon processes as depicted in Fig.\ \ref{singlecomptonpic}. 

In this linear regime of Compton scattering, the parameter $\chi$ is not appropriate to distinguish the importance of quantum effects since it does not give an estimate for the importance of linear quantum effects. As we have mentioned in the introduction, an appropriate parameter to quantify their importance is \cite{RitusRev}
\begin{equation}
\varrho=\frac{(kp)}{m^2}=\frac{\chi}{\xi}=\frac{\omega(\epsilon+p)}{m^2}, \label{linearquantumeffects}
\end{equation}
with the last equality holding in the special coordinate frame introduced in the last paragraph. From Eq.\ (\ref{linearquantumeffects}) it follows that in the regime $\xi \ll 1$, we have $\varrho \gg \chi$, and quantum effects can affect the spectra even for $\chi \ll 1$ as soon as $\varrho \sim 1$; however, in this case they are linear.

In this section we will plot emission spectra only for $\vartheta = \pi$ because at small values of $\xi$ an initially ultrarelativistic electron experiences only a little deviation from its initial propagation direction upon scattering from the laser pulse. 

\begin{figure}[h]
\centering
\includegraphics[width=0.4\linewidth]{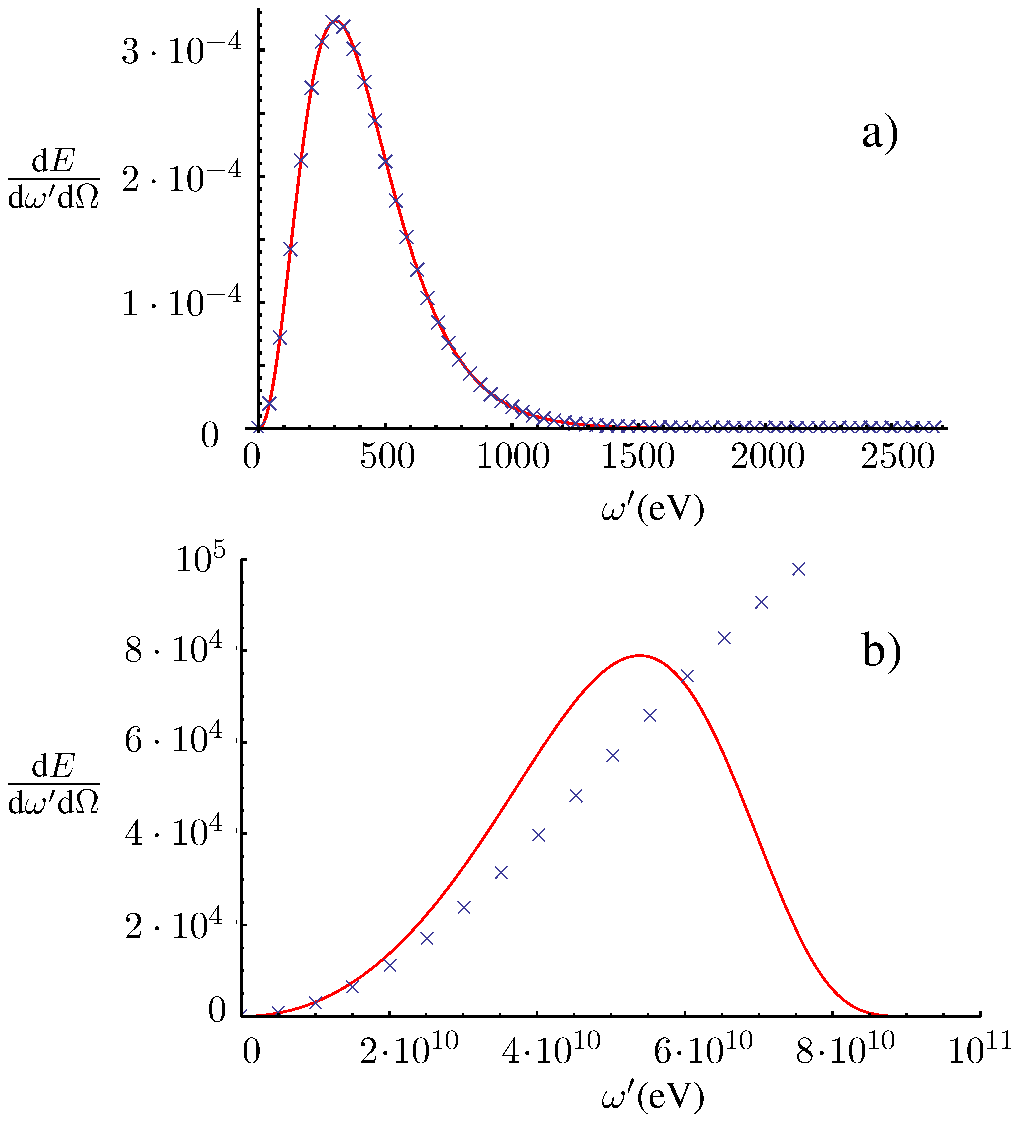}
\caption{(Color online) Energy emission spectra for $\xi=0.05$ and (a) $\gamma=10$ ($\varrho\approx4\times10^{-5}$) and (b) $\gamma=2\times10^5$ ($\varrho\approx0.8$). The blue crosses in both spectra have been obtained by a classical calculation.}
\label{lowintspec}
\end{figure}

Following the procedure outlined in the theory section, we computed perturbative energy spectra. To this purpose, we chose a small nonlinearity parameter $\xi = 0.05$ corresponding to an optical intensity of $I \approx 4.4\times 10^{15}$ W$/$cm$^2$ (here and in the following, the laser photon energy is assumed to be $1\;\text{eV}$). In Fig.\ \ref{lowintspec}(a) we show the classical (crosses) and the quantum (solid red line) emission spectra for an electron with a moderate initial Lorentz factor of $\gamma=10$. The above parameters correspond to a quantum parameter of $\varrho \approx 4\times 10^{-5} \ll 1$, and in fact, we observe that the two spectra are identical. The classical spectrum was obtained by first
solving the Lorentz equation and then plugging the resulting trajectory into the Li\'{e}nard-Wiechert potentials [see Eq.\ (66.9) in \cite{LandauII}]. In Fig.\ \ref{lowintspec}(b), however, we consider an increased Lorentz factor for the incoming electron of $\gamma = 2\times10^5$ leading to $\varrho\approx 0.8$. Thus quantum effects are important in this case. In fact, the spectrum shows two clear differences with respect to the classical spectrum of Fig.\ \ref{lowintspec}(a), only the first of which can be explained classically.
The first difference is that in Fig.\ \ref{lowintspec}(b), considerably higher photon frequencies are emitted. This can be explained through a blueshift of the incoming laser's photons. In fact, since $\xi\ll 1$, the observed peak corresponds to the absorption of one laser photon. In the electron's average rest frame, the central laser photon frequency $\omega^*$ is blueshifted to 
\begin{eqnarray}
\omega^{*\prime}&=&\gamma \omega^*\left(1+\beta_D\right). \label{electronrestfreq}
\end{eqnarray}
In this expression, $\beta_D$ is the electron's drift velocity, that is, the velocity of the reference frame in which the electron would be on average at rest during the scattering process if the scattering laser field were periodic \cite{SalamFais}. In this frame the electron emits a photon with the same frequency as the incident ones. Then the energy of the emitted photon at an angle $\vartheta$ is obtained by going back to the laboratory frame. After some algebra, one finds as the theoretically predicted frequency of the spectrum's maximum
\begin{equation}
\omega'_{\text{Theo}} = \omega^* \frac{1+\beta_D}{1+\beta_D\, \cos(\vartheta)}. \label{omegatheo}
\end{equation}
The maximum of the spectrum in Fig.\ \ref{lowintspec}(a) agrees very well with the frequency $\omega'_{\text{Theo}}\approx300$ eV computed via Eq.\ (\ref{omegatheo}). However, for the parameters of Fig.\ \ref{lowintspec}(b), this equation predicts a peak at $\omega'_{\text{Theo}} \approx 10^{11}$ eV, what is indeed reproduced by the crosses but strongly disagrees with the quantum result. In fact, at very large $\gamma \gg \xi$ and at $\vartheta = \pi$, we have, for Eq.\ (\ref{omegatheo}),
\begin{eqnarray}
 \omega'_{\text{Theo}} &\approx& \omega^* \frac{1+\beta_D}{1-\beta_D} \approx 4\omega^*\, \gamma^2,
\end{eqnarray}
that is, $\omega'_{\text{Theo}}$ grows quadratically with $\gamma$. Now, looking back at Eq.\ (\ref{omgmx}), we see that there exists a maximally allowed emission frequency, which for large $\gamma$ and $\vartheta = \pi$ reads as 
\begin{equation}\omega'_{\text{max}} = m\,\gamma,\end{equation}
which grows linearly with $\gamma$. Thus, by comparing the latter two equalities, we already infer that at some value of $\gamma$, the blueshifted center frequency of the incident laser pulse will exceed the maximally allowed emission frequency. At these values of $\gamma$, the spectra will change their shapes, and the radiation will pile up toward the cutoff energy $\omega'_{\text{max}}$ \cite{Hartemann}. These considerations can be put in a more mathematical form by observing that classically, the spectra at small values of $\xi$ are essentially given by the Fourier transform of the laser field (in this regime, the parameters $\alpha$, $\beta$, and $\zeta$ are proportional to $\omega'$). On the other hand, in the quantum regime, the parameters $\alpha$, $\beta$, and $\zeta$ have a nonlinear dependence on $\omega'$, and they diverge in the limit $\omega'\to\omega'_{\text{max}}$. In this way, the resulting strong oscillations damp the spectrum in the same limit.

\begin{figure}[h]
 \centering
 \includegraphics[width=0.4\linewidth]{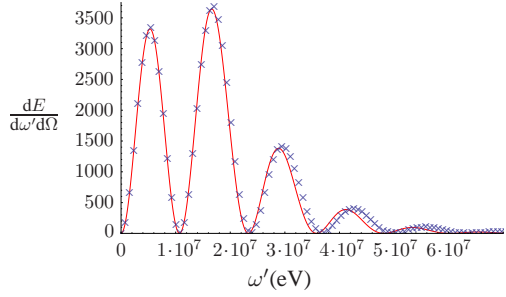}
\caption{(Color online) Energy emission spectrum for $\xi=2$ and $\gamma=2500$ ($\varrho = 10^{-2}$). The crosses have been obtained by a classical calculation.}
\label{destintspec}
\end{figure}

By increasing the value of the nonlinearity parameter to $\xi \gtrsim 1$, new features arise in the energy emission spectra, and if $\varrho\ll 1$, they can also be interpreted classically. In Fig.\ \ref{destintspec} we show the emission spectrum for $\xi = 2$ and an initial $\gamma$ factor of the electron of $\gamma=2500$.

This choice corresponds to a quantum parameter of $\varrho = 10^{-2}$, and the crosses again give the results of a classical computation. Although there are slight discrepancies between classical and quantum results toward higher emission frequencies, which actually hint at upcoming quantum effects, up to around the first minimum, they still match very well. The appearance of several minima and maxima in the spectrum is the main difference to the previous case in which $\xi\ll 1$. The position of the minima can be interpreted both classically and in terms of a quantized photon field. 
The quantum picture is easily established, reminding us that the parameter $\xi$ is interpretable as the average number of photons absorbed by the electron over one Compton wavelength. Then the additional peaks in the spectra can be viewed as multiphoton peaks. The positions of these maxima are the integer multiples of the blueshifted ground frequency predicted by Eq.\ (\ref{omegatheo}). By computing the blueshifted ground frequency $\omega'_{\text{Theo}}$ for $\xi=2$ and $\gamma=2500$, as in Fig.\ \ref{destintspec}, we find 
\begin{equation}
\omega'_{\text{Theo}}\approx6\times 10^6\, \text{eV},
\end{equation}
which indeed agrees well with the first maximum of Fig.\ \ref{destintspec}.

\begin{figure}[h]
\centering
\includegraphics[width=0.5\linewidth]{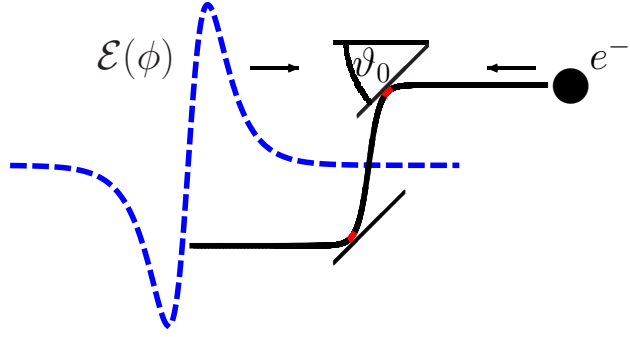}
\caption{(Color online) Classical electron trajectory (solid line) inside a field given by Eq.\ (\ref{elfield}) (dashed blue line).}
 \label{electrontraj}
\end{figure}

Classically the pattern is interpreted as interferences between the field emitted by the electron in different parts of its trajectory. In fact, it can be seen that the electron emits twice into every angle $\vartheta_0$. To clarify this, we show in Fig.\ \ref{electrontraj} the trajectory of an electron (solid curve) moving in a field corresponding to Eq.\ (\ref{elfield}) (dashed blue curve). The two red parts of the trajectory correspond to the emission angle $\vartheta_0$. Furthermore, we note that as visualized in Fig.\ \ref{electrontraj}, a classical electron covers only a certain angle range deviating from its initial direction of propagation when moving in the given electric field. Since its velocity vector will only point into this confined angular region, the electron will also only emit into this region, provided it is ultrarelativistic.

This is the classical analog to the boundary angles found from Eq.\ (\ref{statptcond}). In order to quantitatively investigate the interferences from these two trajectory segments, we employ a classical calculation. Now, classically the differential energy spectrum $\d E/\d \omega'\d \Omega$ can be calculated via, for example, Eq.\ (14.65) in \cite{Jackson}. In this formula we can replace the integral over all times by a sum over the two time instants when the electron emits into direction $\mathbf{n}$. These instants can be distinguished by the phase values $\phi_1 = t_1 - \mathbf{n}\cdot \mathbf{r}(t_1)$ and $\phi_2 = t_2 - \mathbf{n}\cdot \mathbf{r}(t_2)$. We then end up with a formula for the differential energy spectrum
\begin{equation}
\frac{\d E}{\d \omega'\d \Omega} \approx \frac{e^2}{4\,\pi^2} \left|\left\{\frac{\mathbf{n} \times \left[\left( \mathbf{n}-\bm{\beta}_1\right)\times \dot{\bm{\beta}_1} \right]}{\left(1- \bm{\beta}_1 \cdot \mathbf{n} \right)^2} \e^{\imagi\,\omega' \left[t_1 - \mathbf{n}\cdot \mathbf{r}(t_1)\right]}+ \frac{\mathbf{n} \times \left[\left( \mathbf{n}-\bm{\beta}_2\right)\times \dot{\bm{\beta}_2} \right]}{\left(1- \bm{\beta}_2 \cdot \mathbf{n} \right)^2} \e^{\imagi\,\omega' \left[t_2 - \mathbf{n}\cdot \mathbf{r}(t_2)\right]}\right\}\Delta t\right|^2. \label{linwiechsum}
\end{equation}
with $\bm{\beta}_{1/2}:=\bm{\beta}(t_{1/2})$ being the electron's velocity, $\dot{\bm{\beta}}_{1/2}$ its acceleration at the two time instants, and $\Delta t$ the discretized time interval.

The approximation of only two space-time points contributing to the radiation detected under $\vartheta_0$ is better fulfilled the larger the electron's Lorentz factor at the instant of emission is, that is, the narrower its emission cone becomes. In order to evaluate Eq.\ (\ref{linwiechsum}), we note that the points $\phi_1$ and $\phi_2$ by construction satisfy that $\bm{\beta}_1 = \bm{\beta}_2 =:\bm{\beta}$, that the direction of observation is $\mathbf{n}= (\sin(\vartheta_0),0,\cos(\vartheta_0))$, and that, since the electric field from Eq.\ (\ref{elfield}) is antisymmetric around its zero point, the forces acting on the electron at the two points are exactly opposite: $\dot{\bm{\beta}}_1 = - \dot{\bm{\beta}}_2 =: \dot{\bm{\beta}}$. Following these considerations, we have
\begin{eqnarray}
\frac{\d^2 E}{\d \omega'\,\d \Omega} &\approx& \frac{e^2(\Delta t)^2}{4\,\pi^2} \left|\frac{\mathbf{n} \times \left[\left( \mathbf{n}-\bm{\beta}\right)\times \dot{\bm{\beta}} \right]}{\left(1- \bm{\beta} \cdot \mathbf{n} \right)^2}\left\{ \e^{\imagi\,\omega' \left[t_1 - \mathbf{n}\cdot \mathbf{r}(t_1)\right]} - \e^{\imagi\,\omega' \left[t_2 - \mathbf{n}\cdot \mathbf{r}(t_2)\right]}\right\}\right|^2\!\!\!. \label{linwiechsum2}
\end{eqnarray}
So we note that we will have $\d E/\d \omega'\d \Omega = 0$, that is, destructive interference at the frequencies $\omega'_n$, where it holds that
\begin{eqnarray}
\omega'_n = \frac{2\pi\,n}{\left[t_1 - \mathbf{n}\cdot \mathbf{r}(t_1)\right]-\left[t_2 - \mathbf{n}\cdot \mathbf{r}(t_2)\right]}. \label{omegan}
\end{eqnarray}
The method described above of course is not rigorously applicable at the polar angle $\vartheta = \pi$  because many points of the electron's trajectory contribute to the emission in that direction. However, the frequencies at which destructive interference occur can be found by computing the $\omega'_n$ at a polar angle $\vartheta = \pi - \varepsilon$ and then considering the limit $\varepsilon \to 0$. Now, by solving Eq.\ (\ref{statptcond}) for the minimal polar angle where radiation is expected for the choice (\ref{fourpotential}) of $\psi_{\mathcal{A}}$, we find
\begin{equation}
\vartheta_{\text{min}} = 2 \text{ arccot} \left( \frac{\xi}{\gamma \left(1 + \sqrt{1-\frac{1}{\gamma^2}} \right)} \right)\approx\pi - \frac{\xi}{\gamma}.
\end{equation}
Accordingly, by numerically analyzing the electron trajectory, one obtains from Eq. (\ref{omegan}) that
\begin{eqnarray}
\left.\omega'_1\right|_{\vartheta\to\pi}=1.1\times 10^7\, \text{eV},
\end{eqnarray}
matching the observed first frequency of destructive interference from Fig.\ \ref{destintspec} well. It is noteworthy that in the above derivation the ultrashort duration of the scattering laser pulse did not enter. Thus, in the realm of classical electrodynamics, the position of the multiphoton peaks could equally have been recovered in an analysis of a long pulse. In this case, however, the peaks would have been much narrower going in the limit of a monochromatic wave to delta spikes. On the other hand, in Fig.\ \ref{electrontraj} and in all other emission spectra we show, we observe significant broadening of the multiphoton peaks. This is the second essential difference in the scattering of an electron from a long and an ultrashort laser pulse apart from the absence of a mass dressing in the latter case.

\begin{figure}[h]
\centering
\includegraphics[width=0.4\linewidth]{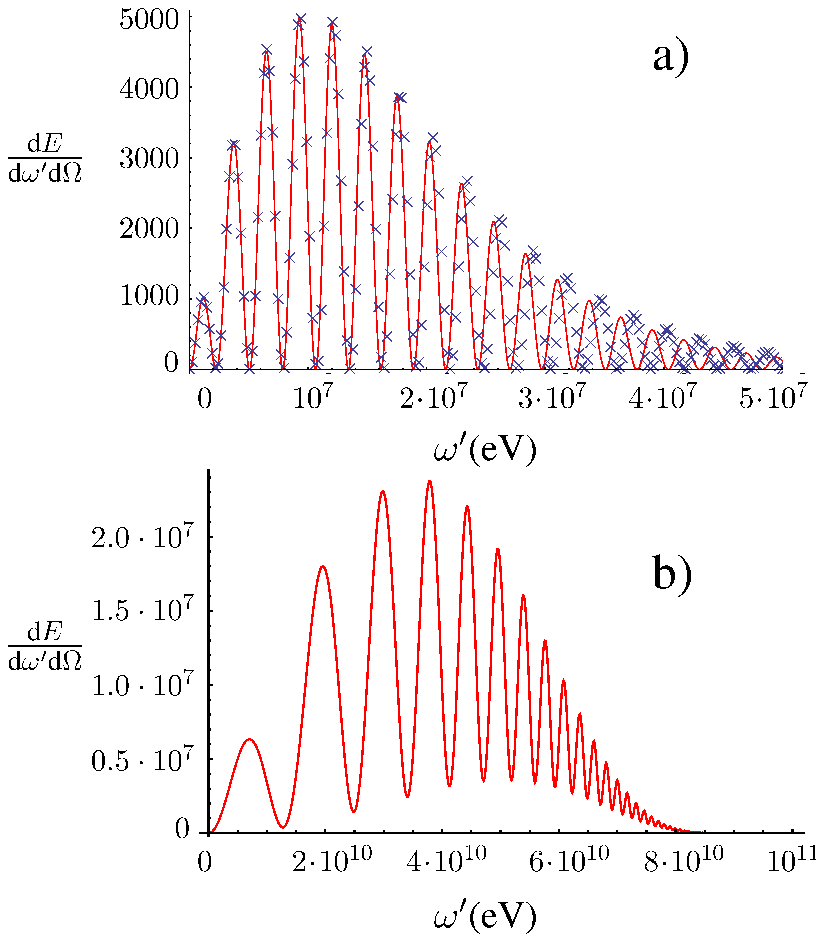}
\caption{(Color online) Energy emission spectra for $\xi=5$ and (a) $\gamma=2500$ ($\varrho\approx2.5\times10^{-2}$) and (b) $\gamma=2\times10^5$ ($\varrho\approx0.8$). The crosses in (a) have been obtained by a classical calculation.}
\label{medintspec}
\end{figure}

It is interesting to note in Eq.\ (\ref{omegan}) that the value of $\omega'_1$ is inversely proportional to the difference $\left(t_1 -t_2\right) -\mathbf{n}\cdot\left[\mathbf{r}(t_1)-\mathbf{r}(t_2)\right]$. Now, at $\vartheta\approx\pi$, it is $\d t-\mathbf{n}\cdot \d\mathbf{r}=(1+\beta_z)\d \phi/(1-\
\beta_z)$. From the analytical solution of the Lorentz equation in a plane wave, it can be seen that at a given $\gamma$, the difference $\left(t_1 -t_2\right) -\mathbf{n}\cdot\left[\mathbf{r}(t_1)-\mathbf{r}(t_2)\right]$ increases at increasing $\xi$. Consequently, the first frequency of destructive interference will be smaller. This assertion is fully consistent with the quantum multiphoton picture. In order to confirm it, we show in Fig.\ \ref{medintspec} the energy emission spectra again at $\gamma=2500$, as in Fig. 8, but with a nonlinearity parameter of $\xi =5$ [Fig.\ \ref{medintspec}(a)].
\\In Fig.\ \ref{medintspec}(a), $\varrho = 2.5\times10^{-2} \ll 1$ and the spectrum is classical (the crosses again give the results of a classical calculation). Comparison with Fig. 8 clearly shows the appearance of many peaks. Computing for the parameters of Fig.\ \ref{medintspec}(a), the frequencies of the first maximum and minimum according to Eq.\ (\ref{omegatheo}) (multiphoton picture) and Eq.\ (\ref{omegan}) (interference picture), respectively, we find
\begin{eqnarray}
 \omega'_{\text{maximum}} &\approx& 1.4\times10^6\,\text{eV}, \nonumber\\
 \omega'_{\text{minimum}} &\approx& 2.2\times10^6\,\text{eV}.
\end{eqnarray}
This is in good agreement with the numerical result. On the other hand, in Fig.\ 10(b), it is $\varrho \approx 0.8$, and quantum effects dominate the spectrum's structure. This is manifest from the fact that the distance between two next peaks decreases as the emission frequency tends to $\omega'_{\text{max}}$.

We finally observe that deviations from Eq.\ (\ref{omegan}) may also arise for not very large $\gamma$ and $\xi$. In this case the electron is not ultrarelativistic, and its emission cone will be relatively wide. The wider the emission cone is, however, the poorer is the assumption that the emission detectable at a certain $\vartheta_0$ originates from only two distinct space points, which was an essential assumption to arrive at Eq.\ (\ref{omegan}).
%
%
\subsection{High intensity regime}
In order to calculate the spectra in the high-intensity regime ($\xi \gg 1$), we substitute the asymptotic expansions (\ref{gnrlasympexp}) into the general formula (\ref{ergspec}). Since the spectra would show numerous maxima and minima, it is clearer to plot only the envelope of the spectra. Furthermore, we show that in fact, in this regime, the parameter $\chi$ is fit to quantify the onset of quantum parameters. Finally, apart from the last two examples (Figs. 14 and 15), we will employ the sine-shaped pulse in Eq. (\ref{fourpotential}).

\begin{figure}[h]
 \centering
 \includegraphics[width=0.4\linewidth]{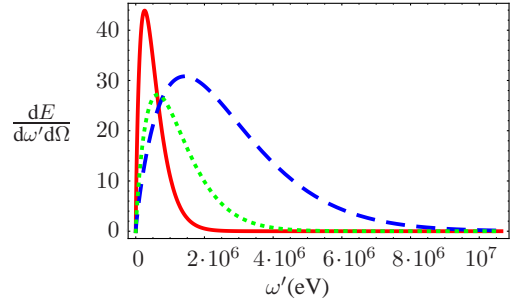}
 \caption{(Color online) Envelopes of the energy emission spectra for $\gamma = \xi =100$ for the emission angles $\vartheta = 127^{\circ}$ (solid red line), $\vartheta = 147^{\circ}$ (dashed blue line), and $\vartheta = 167^{\circ}$ (dotted green line).}
 \label{asymp_spec_chi=0.02}
\end{figure}

We begin by presenting the energy emission spectrum for the parameters $\gamma = \xi = 100$. According to Eq.\ (\ref{chieq}), this corresponds to a quantum nonlinearity parameter of $\chi \approx 0.04$. From Eq.\ (\ref{statptcond}), we find that with this choice, the minimal angle where radiation is expected is $\vartheta_{\text{min}} \approx 127^{\circ}$, and in Fig.\ \ref{asymp_spec_chi=0.02}, we plot this as the smallest scattering angle. .

At $\vartheta_{\text{min}}$, the energy spectrum covers only substantially smaller emission frequencies than at larger angles. The maximal polar observation angle the emission spectrum is plotted for is $\vartheta \approx 167$\textdegree. The reason for this is that, as discussed [Eq.\ (\ref{abkasymp1})], for larger $\vartheta$, the approximation $\alpha \sim \beta \sim \zeta$ does not hold any more. At $\vartheta \approx 164^{\circ}$, for example, we find $\left|\beta/\zeta\right|\approx 9$ and for growing angles, this ratio quickly exceeds 10.

Since for the above parameter choice the quantum nonlinearity parameter $\chi$ is rather small, a classical calculation gives emission spectra comparable to those in Fig. 11. If, according to Eq.\ (\ref{omgmx}), we calculate the maximum frequency that may be emitted from an electron under the given process parameters, we find
\begin{eqnarray}
\omega'_{\text{max}} (\gamma = 100) =
\begin{cases}
63.8\, \text{MeV}&\text{for } \vartheta = 127^{\circ} \\
52.2\, \text{MeV}&\text{for } \vartheta = 167^{\circ}
\end{cases},
\end{eqnarray}
and the maximum frequencies at all other angles in the range $\vartheta \in \left[\vartheta_{\text{min}},\vartheta_{\text{max}}\right]$ lie between these two frequencies. The frequencies actually emitted in the spectra in Fig.\ \ref{asymp_spec_chi=0.02} are approximately 1 order of magnitude smaller than the maximally allowed emission frequency, implying that we are in fact in the classical regime.

\begin{figure}[h]
\centering
\includegraphics[width=0.4\linewidth]{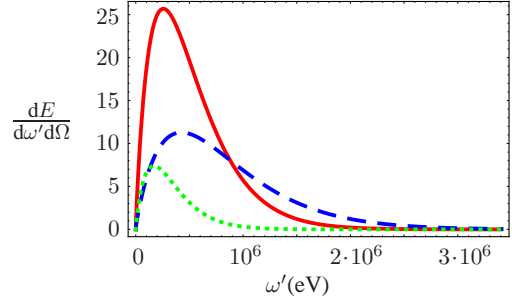}
\caption{(Color online) Envelopes of the energy emission spectra for $\gamma = 50$ and $\xi=100$ for the emission angles $\vartheta = 91^{\circ}$ (solid red line), $\vartheta = 122^{\circ}$ (dashed blue line), and $\vartheta = 154^{\circ}$ (dotted green line).}
 \label{asymp_spec_chi=0.01}
\end{figure}

In order to observe backscattering, that is, \ $\vartheta_{\text{min}} \leq 90^{\circ}$, we have to consider $\xi \geq 2\gamma$. We thus consider a lower value of $\gamma$ and plot the emission spectra in the case of $\xi = 100$ and $\gamma = 50$, such that we expect emission into a large polar angle regime. The quantum parameter in this scenario is $\chi \approx 0.02$, and the resulting spectra are shown in Fig. \ref{asymp_spec_chi=0.01}.

The general shape of these spectra is similar to those previously shown in Fig.\ \ref{asymp_spec_chi=0.02}, albeit with different axis scales. The emission range indeed is found to extend down to emission angles close to $\vartheta = 90^{\circ}$. Due to its lower initial energy as compared to the laser intensity, the deviation of the electron's direction of propagation from its initial orientation is increased. We have ensured that the emission is strongly suppressed at $\vartheta <90^{\circ}$.

\begin{figure}[h]
 \centering
 \includegraphics[width=0.4\linewidth]{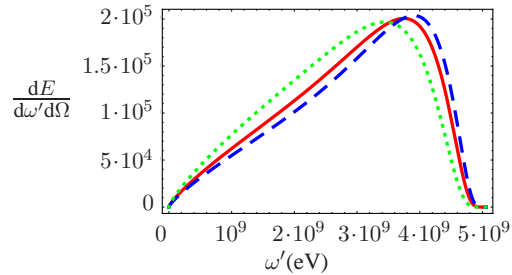}
\caption{(Color online) Envelopes of the energy emission spectra for $\gamma =10^4$ and $\xi =100$ for the emission angles $\vartheta = 179.5^{\circ}$ (solid red line), $\vartheta = 179.6^{\circ}$ (dashed blue line), and $\vartheta = 179.8^{\circ}$ (dotted green line).}
 \label{asymp_spec_chi=2}
\end{figure}

In order to study a case where quantum effects become important, we choose $\xi = 100$ and $\gamma = 10^4$, which gives a quantum parameter of approximately $\chi \approx 4$. The resulting spectra are presented in Fig.\ \ref{asymp_spec_chi=2}.

In the spectra plotted we find clear differences with respect to those in Figs.\ \ref{asymp_spec_chi=0.02} and \ref{asymp_spec_chi=0.01}. Not only do the scales of emitted frequencies and of the spectra's amplitudes differ by several orders of magnitude, but also the shapes of the spectra look distinctly different. So in Fig.\ \ref{asymp_spec_chi=2}, we find a fast decay of the emitted energy when the emitted frequencies approach the maximally emitted photon energy, while in Fig.\ \ref{asymp_spec_chi=0.02}, the spectra exhibited a much more moderate decrease. This again is evidence for the effect of kinematic pile up as described in section \ref{classresultssec}. Computing again the maximally allowed emitted photon frequency according to Eq.\ (\ref{omgmx}), we find that
\begin{eqnarray}
\omega'_{\text{max}} (\gamma = 10^4,\vartheta = 179.5^{\circ}) =
5.11\times10^9\,\text{eV},
\end{eqnarray}
which is close to the frequency where the fast drop off in Fig.\ \ref{asymp_spec_chi=2} occurs. So we conclude that in the regime $\xi \gg 1$, as soon as the quantum parameter $\chi$ becomes of order unity, the emitted photons will approach their maximally allowed frequencies. This interpretation of the spectra's distortions found in Fig.\ \ref{asymp_spec_chi=2} clearly hints at quantum effects which, due to energy-momentum conservation, prevent the emission of higher energetic photons.

\begin{figure}[h]
 \centering
 \includegraphics[width=0.8\linewidth]{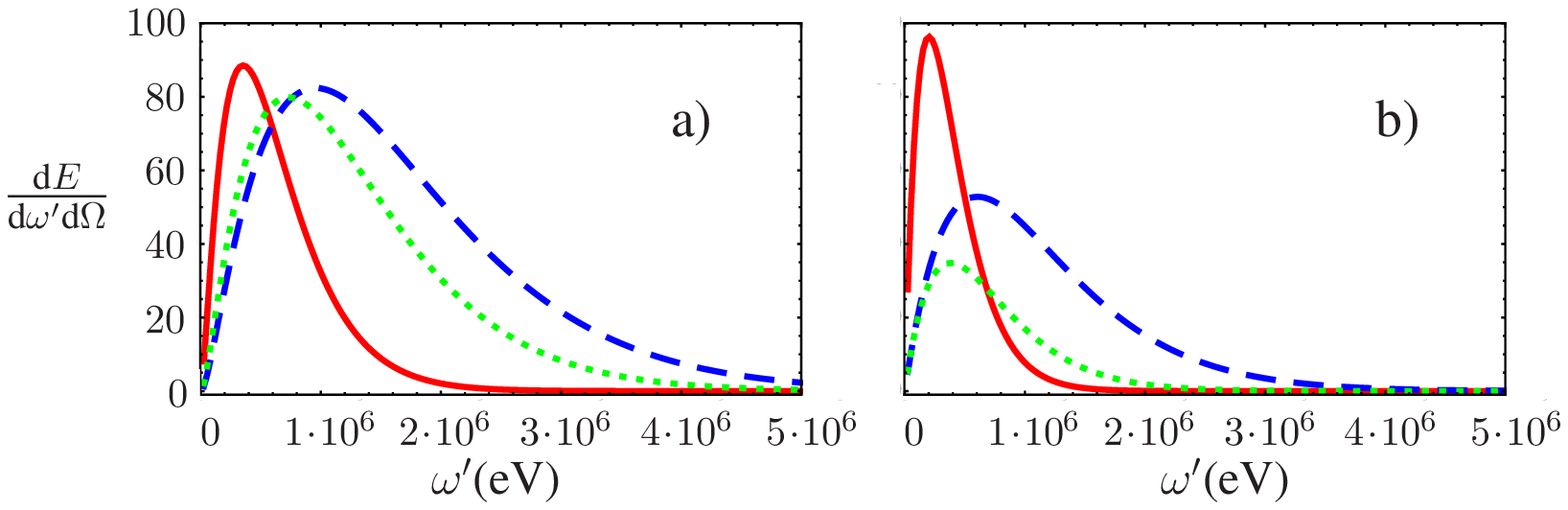}
\caption{(Color online) (a) Envelopes of the energy emission spectra for $\gamma = \xi =100$ at the emission angles $\vartheta = 152^{\circ}$ (solid red line), $\vartheta = 160^{\circ}$ (dashed blue line), and $\vartheta = 167^{\circ}$ (dotted green line).\\(b) Envelopes of the emission spectra for $\gamma = 50$, $\xi = 200$ at the emission angles $\vartheta = 90^{\circ}$ (solid red line), $\vartheta = 110^{\circ}$ (dashed blue line), and $\vartheta = 131^{\circ}$ (dotted green line).}
 \label{asymp_spec_CEP_chi=0.04}
\end{figure}

Up to now, all emission spectra were obtained for the choice of $\psi_{\mathcal{A}}$ corresponding to a sine-shaped electric field [Eq.\ (\ref{fourpotential})]. To investigate the scattering off a cosine-shaped pulse, we computed emission spectra for the choice (\ref{fourpotential1}). For this choice from Eq.\ (\ref{statptcond}) we do not only expect emission into the azimuthal angle regime around $\varphi = 0$ but also to $\varphi = \pi$. We begin in Fig.\ \ref{asymp_spec_CEP_chi=0.04}(a) by showing the spectra for $\xi = \gamma = 100$, which makes it easy to compare the results to those shown in Fig.\ \ref{asymp_spec_chi=0.02}. We have ensured that the spectra at $\varphi=\pi$ are the same as those at $\varphi=0$, and we show only the case $\varphi=0$.

We note that in Eq.\ (\ref{statptcond}), the values $\psi_{\mathcal{A},\text{min/max}}$ are changed. Thus the angular range where emission is predicted changes with respect to the spectra in Fig.\ \ref{asymp_spec_chi=0.02} for an unchanged ratio $\xi /\gamma$. In Fig.\ \ref{asymp_spec_CEP_chi=0.04}, we accordingly plotted the spectrum at $\vartheta = 152^{\circ}$ as the smallest polar angle because it is found from Eq.\ (\ref{statptcond}) that the emission extends only to larger angles.

In Fig.\ \ref{asymp_spec_CEP_chi=0.04}(b), we show the emission spectra for $\gamma = 50$ and $\xi =200$. In these spectra we find the minimal emission angle to be $\vartheta_{\text{min}}=90^{\circ}$. So it becomes obvious that for the choice (\ref{fourpotential1}), in order to observe radiation in the space region $\vartheta_{\text{min}}<90^{\circ}$, a nonlinearity parameter $\xi \ge 4\gamma$ must be employed. This is a clear difference to the sine-shaped pulse, where for backscattering to occur, it had to be satisfied that $\xi \geq 2\gamma$.

\begin{figure}[h]
 \centering
 \includegraphics[width=0.4\linewidth]{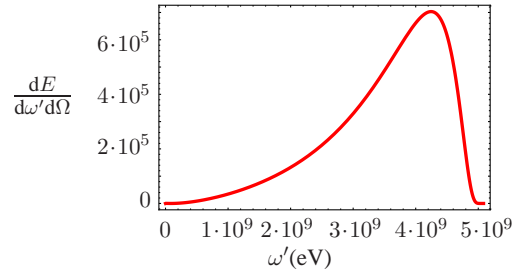}
\caption{(Color online) Envelope of the energy emission spectrum for $\gamma = 10^4$ and $\xi =100$ for the emission angle $\vartheta = 179.8^{\circ}$.}
 \label{asymp_spec_chi=4}
\end{figure}

Finally we show a spectrum for the cosine-shaped pulse where quantum effects become important. For this purpose we choose the parameter set $\xi = 100$ and $\gamma=10^4$, corresponding to a quantum parameter $\chi = 4$. The resulting spectrum is shown in Fig.\ \ref{asymp_spec_chi=4}.

As for the case of the sine-shaped pulse, we observe in this spectrum the typical quantum pile-up of the radiation toward $\omega'_{\text{max}}$.
%
%
\section{Conclusions}\label{conclusionssec}
As we mentioned in the introduction there has already been a lot of work on Thomson and Compton scattering from monochromatic \cite{KibbleBrown,NikiRitus,BaierMilstein,RitusRev} or long-pulsed \cite{NarozhnyiFofanov1} lasers fields. In the present article we have considered the opposite situation in which the pulse contains only one or a few cycles. Comparing our results to previous work, we find some agreement as well as some differences. We obtain analogous qualitative behavior in our emission spectra concerning the rise of nonlinear quantum effects in the scattering process. Specifically we could show that in the regime $\xi \gg 1$, the parameter $\chi$ also in the case of ultrashort laser pulses is suitable to characterize the onset of quantum effects. However, we could unveil two major differences between these earlier treatments and our analysis. 

First of all, we found the momentum conserving $\delta$ functions contained in the transition matrix element to differ from those obtained for scattering off monochromatic laser waves or long pulses. We interpreted this as the absence of a dressed mass effect in particular for ultrashort pulses containing only one laser cycle. Additionally we pointed out that this prediction could be tested by not only detecting the photons emitted in the scattering process but also the scattered electrons. Then, by measuring the momenta of the outgoing particles, one could judge if the electron propagated with a dressed mass inside the pulse or not. Second, we found a different structure of the energy emission spectra. Even though the multiphoton peaks are still present, they are significantly broadened with respect to the case of a monochromatic laser beam. This can be understood such that for ultrashort durations a laser pulse can no longer be viewed to be composed of many photons with the same energy. These two features are the effects we find to be expected when considering ultrashort pulses in an electron-laser scattering event.

Moreover we have shown analytically that, when classical electrodynamics apply, a classical interference picture and a quantum mechanical multiphoton description both consistently describe the position of the minima and maxima in the emission spectra. Finally, by means of a careful application of the saddle-point method, we have been able to calculate analytically the spectra at large values of $\xi$ also for ultrashort laser pulses.

\end{document}